\documentclass[aps,prd,twocolumn,superscriptaddress,showpacs,floatfix]{revtex4}
\usepackage{bm, natbib, hyperref, graphicx, amsmath}

\newcommand{\lok}{\left(}
\newcommand{\rok}{\right)}
\newcommand{\lensa}{\left<}
\newcommand{\rensa}{\right>}
\newcommand{\lkw}{\left[}
\newcommand{\rkw}{\right]}
\newcommand{\beq}{\begin{equation}}
\newcommand{\eeq}{\end{equation}}
\newcommand{\beqa}{\begin{eqnarray}}
\newcommand{\eeqa}{\end{eqnarray}}
\newcommand{\xie}{\xi^{\ee}}
\newcommand{\pkk}{ P_{\kappa} }
\newcommand{\masst}{ M_{\rm ap} }
\newcommand{\massr}{ M_{\times} }
\newcommand{\dd}{\delta}
\newcommand{\DD}{\Delta}
\newcommand{\dD}{\delta_{\rm D}}
\newcommand{\gt}{\gamma_{+}}
\newcommand{\gr}{\gamma_{\times}}
\newcommand{\gam}{\gamma}
\newcommand{\gx}{\gamma_{1}}
\newcommand{\gy}{\gamma_{2}}
\newcommand{\kk}{\kappa}

\newcommand{\tht}{\bm{\theta}}
\newcommand{\bphi}{\bm{\phi}}

\newcommand{\vth}{{\bm \theta}}

\newcommand{\ee}{\epsilon}
\newcommand{\oo}{\omega}

\newcommand{\xigam}[1]{\xi^{\gam}_{#1}}
\newcommand{\xid}[1]{\xi^{d}_{#1}}
\newcommand{\vd}{{\bm d}}

\newcommand{\vg}{{\bm \gamma}}
\newcommand{\arcdeg}{\ensuremath{^\circ}}
\newcommand{\arcmin}{\ensuremath{'}}

\begin{document}

\title{Inhomogeneous systematic signals in cosmic shear observations}
\author{Jacek Guzik}
\email{guzik@astro.upenn.edu}
\affiliation{Department of Physics and Astronomy, University of Pennsylvania, Philadelphia, PA 19104, U.S.A.}
\affiliation{Astronomical Observatory, Jagiellonian University, Orla 171, 30-244 Krak\'ow, Poland}
\author{Gary Bernstein}
\email{garyb@physics.upenn.edu}
\affiliation{Department of Physics and Astronomy, University of Pennsylvania, Philadelphia, PA 19104, U.S.A.}
%\date{Received 10 May 2005; published 5 August 2005}

\begin{abstract}
We calculate the systematic errors in the weak gravitational lensing
power spectrum which would be caused by spatially varying calibration
(i.e., multiplicative) errors, such as might arise from uncorrected
seeing or extinction variations.
The systematic error is fully described by the angular two-point
correlation function of the systematic 
in the case of the 2D lensing that we consider here.  
We investigate three specific
cases: Gaussian, ``patchy'' and exponential correlation functions. 
In order to keep systematic errors below
statistical errors in future \emph{LSST}-like surveys, the spatial
variation of calibration should not exceed $3\%$ rms.
This conclusion is independently true for all forms of correlation
function we consider.  The relative size the E- and B-mode power
spectrum errors does, however, depend upon the form of the correlation
function, indicating that one cannot repair the E-mode power spectrum
systematics by means of the B-mode measurements.  
\end{abstract}


\maketitle

\section{Motivation}

The power spectrum of the weak gravitational lensing distortions of
background galaxies is quite directly related to the power spectrum of
intervening matter \citep{1991ApJ...380....1M, 1992ApJ...388..272K}.  
The weak lensing (WL) power
spectrum depends upon the linear and non-linear rates of growth of
structure since recombination, and upon the redshift-distance relation
produced by the expansion.  These dependences, plus the
straightforward theoretical framework, make WL a very attractive tool
for the constraint of the post-recombination Universe, e.g. dark
energy.  Current 5--10\% measurements of the WL power spectrum  have
already begun to place interesting 
constraints \citep{2005astro.ph..2243J}, and very much larger-scale projects are
planned to reduce the statistical errors on the WL signal to 1 part in
$10^{3}$ or lower.

To reap the benefits of these large surveys, systematic errors must be
well below the small expected statistical errors.  WL measurements are
subtle and difficult compared to most astronomical data analyses.
There are no ``standard lenses'' on the sky, so calibration of the WL
shear data is a significant worry.  The finite point spread function (PSF) width tends to
circularize the appearance of background galaxies, squelching the WL
shear signal.  This must be corrected analytically, and any errors in
this process, or inaccuracies in the estimate of the PSF size, will
lead to calibration errors.  \citet{2005astro.ph..6030H} investigate the effect of
overall mean shear calibration errors on cosmological parameter
estimation.  It is also likely, however, that there will be spatially
varying calibration errors that are larger than the error in the mean
calibration.  For example, as the PSF size $\sigma_\star$ varies during a
ground-based survey, the resolution parameter $R\equiv
1 - \frac{\sigma^2_\star}{\sigma^2_g}$ of the galaxies will vary (here $\sigma_g$ is
the angular size of a target galaxy).  If we fail to track this
variation properly, the inferred shear will be modulated by a factor
$(1 - \delta R(\tht)/R(\tht))$ \citep{2002AJ....123..583B}. 
In this paper we
calculate the effect of spatially varying calibration errors upon
the measured power spectrum, and determine criteria on these
systematic errors which will have to be met if they are to be made
negligible in future surveys.

Other spatially varying errors
could arise from photometric errors or Galactic extinction 
which will modulate the effective depth of the survey.
Both effects may lead to errors in the (photometric)
redshift estimation and source galaxy distribution, see e.g.  
\citep{2003AJ....125.1014J}, which would in turn lead to local modulation
of the observed shear.  While a modulation of redshift depth is not
strictly equivalent to a multiplicative modulation of shear, our
calculations will still permit an estimate of the level at which depth
modulations become significant.

The lensing shear of sources at some redshift $z_s$ is a 2-component
tensor function of the angular variable $\tht$.  As reviewed
briefly below, the shear field can be divided into ``E'' and ``B''
modes corresponding to curl-free and divergence-free deflections, with
corresponding power spectra $P_E(l)$ and $P_B(l)$ for 
stationary isotropic fields.
Gravitational deflections, being derived from the scalar potential,
will produce only E-mode power in the weak limit.  It is therefore
$P_E(l)$ that will be used for cosmological constraints, with the B mode
serving as the ``canary in the coal mine'' to alert us to
potential non-gravitational sources of systematic error.  A spatially
varying scalar calibration factor will alter the amplitude of the
E-mode power and convert some into B-mode power.  In \S\ref{analytics}
we quantify this effect in terms of the 2-point statistics of the calibration
systematic.  In \S\ref{models} we present solutions using several
models for the systematic error, and show how the deleterious effects
are in general determined just by the rms amplitude and characteristic
angular scale of the calibration errors.  \S\ref{results} summarizes
the results and the requirements upon future surveys that can be
derived from these results.

Some related calculations exist in the literature.  \citet{2002A&A...389..729S}
investigate the B-mode signal that is created by inhomogeneous source
distributions, using a formalism similar to that employed here.  We
note that the inhomogeneous-source effect can be avoided by
considering only cross-correlations between source bins that are
disjoint in redshift.  The calibration inhomogeneity that we analyze
here will likely not be so easily avoided.
\citet{2004ApJ...613L...1V} conduct a numerical test of calibration
inhomogeneity by modulating the shear seen in a ray-tracing
simulation, and then calculating the resultant power spectra.  We will
test our analytic results against their simulated data.

A rough target for calibration systematics is that their effect on
$P_E(l)$ be smaller than the expected statistical errors.  For a
single-screen lens analysis, the uncertainty in $P_E(l)$ averaged over
an interval $\Delta\ln l=1$ is $l^{-1}f_{\rm sky}^{-1/2}P_E(l)$ in the
sample-variance limit.  So for ambitious surveys with $f_{\rm sky}\gtrsim 0.5$, 
the power spectrum statistical errors are $\approx
1$ part in $10^3$ at $l=1000$.  At higher $l$, the uncertainties due
to shape noise and inaccuracies in the non-linear clustering theory
will become important.  
The tolerances may be tighter when
one examines the impact of power-spectrum tomography rather than just
a single power spectrum.  So a good goal is to have the
calibration-induced error $\Delta P_E(l)$ be $\le 10^{-4} P_E(l)$. 

\section{Shear field decomposition into E and B modes}
Decomposition of a spin-2 field, such as shear or the Stokes
parameters, into curl-free and divergence-free part was suggested 
to be useful for weak gravitational lensing studies by \citet{1996astro.ph..9149S}, and 
for cosmic microwave background (CMB) polarization  
by \citet{1997PhRvD..55.7368K} and \citet{1997PhRvD..55.1830Z}. \citet{2002ApJ...568...20C} and 
\citet{2002A&A...389..729S} study its use in revealing non-gravitational signals in weak lensing surveys. 
We briefly review the decomposition of the shear field into independent
E and B modes, following the notation of \citet{2002A&A...389..729S}.

The gravitational lens equation in the one-screen approximation relates the
detected direction $\bm{\theta}$ of photons on the sky 
to the (unobservable) direction $\bm{\beta}$ of photons emitted by a source:
$\bm{\Delta} = \tht - \bm{\beta}$, where $\bm{\Delta}$ is the deflection angle 
scaled by a factor depending on the angular 
diameter distances in the observer-lens-source system 
\citep{2001PhR...340..291B}. 
The gradient of the deflection field, being a tensor of rank two, is
usually decomposed locally into the trace,  
symmetric traceless part, and antisymmetric part as follows $\Delta_{i,j} = \kappa \, \dd_{ij} + \gamma_{ij} + \oo \, \ee_{ij}$,
where the shear tensor $\gamma_{ij}$ is symmetric and $\ee_{ij}$ is
the Levi-Civita symbol in two dimensions. We denote partial
derivatives with respect to directions in the tangent plane on the sky in a standard fashion by a comma.
Thus we may express the convergence $\kk$ and the rotation $\oo$ as linear combinations of derivatives of the deflection 
angle: $ 2\kk  = \DD_{1,1} + \DD_{2,2}$, $2\oo  =  \DD_{1,2} - \DD_{2,1}$.
Also, the shear components $(\gamma_1, \gamma_2)$, defined as 
$\gx \equiv  \gamma_{11} = -\gamma_{22}$, $\gy \equiv \gamma_{12} = \gamma_{21}$, may be written as
$2\gx  =  \DD_{1,1} - \DD_{2,2}$,  $2\gy  =  \DD_{1,2} + \DD_{2,1}$. Moreover, 
we can write the deflection field as a sum of curl free and divergence free parts 
$\bm{\Delta} = \bm{\Delta}_{+} + \bm{\Delta}_{\times}$ which can be expressed as the gradient 
of a scalar potential $\phi_{+}$ and the curl of a pseudoscalar potential $\phi_{\times}$ respectively \citep{1996astro.ph..9149S}.
We designate as ``E-mode'' the curl-free deflection $\bm{\Delta}_{+}$, which
resembles an electric field pattern, and can be due to the mass 
distribution. It produces the tangential shear pattern $\gt$ \citep{1996astro.ph..9149S}.
On the other hand, the divergence-free ``B-mode''
deflection $\bm{\Delta}_{\times}$ resembles a magnetic field pattern. This mode reveals in measurements as a ``radial'' 
shear $\gr$ (i.e., $\gt$ rotated by $45\arcdeg$) and it cannot be generated by lensing in the single-screen approximation.
The potentials $\phi_{+}$  and $\phi_{\times}$ are closely related to the convergence $\kk$ and rotation $\oo$ via the Poisson equation, 
$\nabla^2 \phi_{+} = 2 \, \kk$ and $\nabla^2 \phi_{\times} = 2 \, \oo$. 
Since the single-screen approximation is thought to be valid
in cosmological situations \citep{2003astro.ph..5089V}, gravitational
lensing information is confined to the E mode while the B mode
should be zero. 
Thus the presence of non-zero B mode would be due to breaking of the
single-screen approximation or, more importantly, to a
variety of processes not related 
directly to lensing, such as measurement calibration errors \citep{2003MNRAS.343..459H, 2005A&A...429...75V, 2004astro.ph.12234J}, 
clustering of source galaxies \citep{2002A&A...389..729S}, or their intrinsic alignments  \citep{2003MNRAS.339..711H}. 
For a more thorough discussion of E/B-mode decomposition see \citet{2002ApJ...568...20C}.

In order to quantify the contribution of systematic uncertainties to
the E and B mode power spectra 
we introduce a pair of two-point correlation functions $\xigam{+}(\theta)$ 
and $\xigam{-}(\theta)$, following \citet{2002A&A...389..729S}.
They are linear combinations of correlation functions of  
E and B components of the shear,
defined for each pair of galaxies with respect to the preferred coordinate system 
in which their positions are $\tht_1=(0, 0)$ and $\tht_2=(\theta, 0)$:
\beq
	\xigam{\pm}(\theta) = \lensa \gamma_1(\tht_1) \, \gamma_1(\tht_2) \rensa \pm \lensa \gamma_2(\tht_1) \, \gamma_2(\tht_2) \rensa. \\
	\label{corrpm}
\eeq
Moreover, the correlation functions (\ref{corrpm}) can be expressed as
follows in terms of E and B-mode power spectra, $P_E(l)$ and $P_B(l)$, defined 
as $\lensa \kappa (\bm{l}) \kappa (\bm{l'}) \rensa \equiv \lok 2\pi \rok^2 \dD(\bm{l} + \bm{l'}) \, P_E(l)$ and
$\lensa \omega (\bm{l}) \omega (\bm{l'}) \rensa \equiv \lok 2\pi \rok^2 \dD(\bm{l} + \bm{l'}) \, P_B(l)$:
\beq
	\xigam{\pm}(\theta) = \frac{1}{2\pi} \int_0^{\infty} dl \, l \, \lok P_E(l) \pm P_B(l) \rok \, J_{0,4}(l\theta). 
	\label{xi_pm}
\eeq
We can invert those relations and obtain power spectra expressed in terms of the correlation functions
[in what follows we use a convention that upper sign in the sum on the right 
hand side refers to E-mode power spectrum, lower to B-mode]:
\beq
	P^{\gam}_{E,B} (l)  =  \pi \int_0^{\infty} d\theta \, \theta \, \lkw \xigam{+} (\theta) \, J_{0} (l\theta) 
	\pm \xigam{-}(\theta) \, J_{4} (l\theta) \rkw.
	\label{pseb}
\eeq
We do not consider cross-power spectrum of E and B modes as it will vanish due to parity conservation
\citep{2002A&A...389..729S}.

\section{Effect of systematics on E/B power spectra}
\label{analytics}

Ideally, we would like to measure the shear field $\bm{\gam}(\tht)$ directly.
What we observe, however, is the coherent ellipticity induced on an
ensemble of galaxies, which (a) is defined by the distortion
$\bm{g}=\bm{\gam}/(1-\kappa)$; (b) is imparted on galaxies that are not
intrinsically circular, and (c) are viewed through a finite
point-spread function (PSF).  
The measured shear field $\bm{d}(\tht)$ will in practice be modulated
or contaminated by various 
observational effects \citep{2003astro.ph..5089V}. 
Although techniques for shear extraction from galaxy images have been
extensively developed and tested \citep{1995ApJ...449..460K,
  2002AJ....123..583B, 2003MNRAS.343..459H}, there remain
imperfections which 
can be detrimental to precision cosmology.

Throughout the paper we assume that the observed field is related to
the true shear field by a position-dependent 
multiplicative scalar factor $1+\ee(\tht)$ such as will result from a misestimation 
of the ``resolution'' \citep{2002AJ....123..583B} or ``shear polarizability'' \citep{1995ApJ...449..460K}. 
The systematic field
$\ee(\tht)$ is a random field assumed to have
zero mean and described to the lowest interesting order by the two-point correlation function. 
Thus we express the observed field in terms of the shear and the systematics fields as
\beq
	\bm{d}(\tht) = (1+\ee(\tht)) \, \bm{\gam}(\tht).
	\label{basic}
\eeq
This relation is local in real space, so it is going to couple modes of the shear field 
in the Fourier space, {\it i.e.} have some non-local effect
on the relevant power spectra.
We assume that the shear field $\bm{\gam}$ due to massive structures in the Universe
is uncorrelated with systematics field $\ee(\tht)$, which is a
Galactic or instrumental foreground. The observed
two-point correlation function $\xid{}(\theta)$ can in this case be written as 
\beqa
	\xid{}(\theta) & \equiv & \left<\vd(\bphi) \, \vd(\bphi + \vth)\right> \\ 
	             & =& (1+ \left<\ee(\bphi) \, \ee(\bphi + \vth)\right>) \, \left<\vg(\bphi) \, \vg(\bphi + \vth)\right> \\
	             & =  & \lok 1 + \xie(\theta) \rok \, \xigam{}(\theta).
		     \label{corr1}
\eeqa
We have introduced two-point correlation functions $\xigam{}(\theta)$ for the shear field and $\xie(\theta)$ for the systematics field. 
For simplicity we will assume that the systematics field $\ee(\vth)$
is homogeneous and isotropic.  In practice the assumption of isoptropy
is not restrictive, as the effects of an anisotropic systematic could
be approximated to first order by considering the
azimuthally averaged correlation function.

Correlation functions for the distortion field, $\xid{+}(\theta)$ and $\xid{-}(\theta)$,  
may be expressed as products of the correlation functions for the shear and systematics 
$\xid{\pm}(\theta) = \lok 1 + \xie(\theta) \rok \xigam{\pm}(\theta)$ which follow from eqs. (\ref{corrpm}) and (\ref{corr1}).
We can rewrite eq. (\ref{pseb}) in terms of the distortion instead of the shear
and then account for systematic signals $\ee(\vth)$. 
We split the observed E and B mode power spectra $P^{d}_{E,B}(l)$ into two 
contributions as follows
\beq
	P^{d}_{E,B}(l) = P^{\gam}_{E,B}(l) + \Delta P^{\ee}_{E,B}(l),
	\label{ps_sys1}
\eeq
where the term $P^{\gam}_{E,B}(l)$ is E mode (B mode) power spectrum of the shear 
and $\Delta P^{\ee}_{E,B}(l)$ represents the contributions to the E mode (B mode) power due to systematic signals. 
We focus on these error terms in the remainder of the paper.
Using eqs. (\ref{pseb}) and (\ref{ps_sys1}) they can be written as
\beq
	\Delta P^{\ee}_{E,B}(l) =  \pi \int_0^{\infty} d\theta  \theta  \xie(\theta) \lkw \xigam{+} (\theta) J_0 (l\theta) 
	\pm \xigam{-}(\theta) J_4 (l\theta) \rkw.
	\label{peb}
\eeq
We assume that the shear correlation functions $\xigam{\pm}$
receive contribution from E mode only, i.e. $P^{\gam}_{E}(l) =
\pkk(l)$, $P^{\gam}_{B}(l) \equiv 0$, since 
B-mode cosmological contributions are expected to be a few orders
of magnitude smaller on scales $>1\arcmin$ \citep{2002A&A...389..729S}.
The systematic errors $\Delta P_{E,B} (l)$ to E and B mode power
spectra can be written as integrals
over the convergence power spectrum $\pkk(l)$
with a window function $W_{E,B}(l,q)$:
\beq
	\Delta P_{E,B} (l) = \int_0^{\infty} dq \, q \, \pkk(q) \, W_{E,B}(l,q), 
	\label{ps_sys2}
\eeq
where the window function depends solely on the correlation function
$\xie$ of the systematic modulation, and is given by
\beq
	W_{E,B} = \frac{1}{2} \int_0^{\infty} d\theta \theta \xie(\theta) 
	\lkw J_0 (l\theta) J_0 (q\theta) \pm J_4(l\theta)  J_4 (q\theta) \rkw. 
	\label{win}
\eeq
In the limit of a systematic that is completely correlated across the entire
observation, {\it i.e.} a constant calibration error, we have
$\xie(\theta) = \Sigma^2$,  where $\Sigma^2$ is the variance of the
calibration error.  In this limit
we obtain  $W_{E}(l,q) = \Sigma^2 \, q^{-1} \dD (l-q)$ and $W_{B}(l,q) \equiv 0$, where we have used 
an integral relation for the Bessel functions $\int_0^{\infty} d\theta
\, \theta \, J_n(l\,\theta) \, J_n(q\, \theta) 
= q^{-1} \, \dD(q-l) $ \citep{AS}.  
Thus the error contributions to E/B power spectra are $\Delta P_{E} (l) = \Sigma^2 \pkk(l)$ and
$\Delta P_{B} (l) = 0$  in this case, and there is no conversion of E
power to B power, as expected.

Numerical simulations of calibration inhomogeneity in \citep{2004ApJ...613L...1V}
are presented in terms of the 
aperture mass statistics $\masst(R)$ and $\massr(R)$ with compensated filter defined in
\citep{1998MNRAS.296..873S, 1999A&A...345...17B}. 
We produce analytic predictions for inhomogeneous calibration errors for
comparison with the numerical results of \citep{2004ApJ...613L...1V} using the same filter as they did. 

\section{Modeling of systematics}
\label{models}

We  consider several potentially useful models of the correlation function of the systematic signal $\xie(\theta)$
and we examine the dependence of E and B mode power spectra
(\ref{ps_sys2}) on the characteristics of $\xie(\theta)$.  The
correlation functions considered here are analytically tractable and
able to describe a wide variety of random processes leading to systematic signals.
Each correlation function considered here is assumed to describe a
stationary, isotropic random field. 
We assume that the systematic field has a finite variance $\Sigma^2$.
Equation (\ref{win}) shows that 
$\Delta P_{E,B}(l) \propto \Sigma^2$ if $\Sigma^2$ is a prefactor to some
otherwise fixed functional form for $\xie$.  
Moreover, a correlation function in 2-D has to be bounded from below by a global minimum value of the Bessel function  
$J_0(x)$.  This condition is met by our models because they are assumed to be non-negative \citep{Ripley81}.

We also introduce for each correlation function a
characteristic scale $R_{1/2}$ where the correlation 
function drops to 50\% of its zero-lag value $\Sigma^2$.

\subsection{Gaussian family}
\label{gauss-type}

As a first model let us consider correlation function having a Gaussian 
shape with characteristic scale $\theta_0$  
\beq
	\xi^{\ee} (\theta) = \Sigma^2 \, e^{-\frac{\theta^2}{2\theta_0^2}}.
	\label{gauss}
\eeq
We have $R_{1/2}^2 = 2 \, \theta_0^2 \, \ln 2$ for the Gaussian.   
The Gaussian is chosen because they are usually easy to handle
analytically.
In this case an integral over scale $\theta$ in  eq. (\ref{win}) can be 
done analytically \citep{GR} and we obtain the following window function
\beq
	W_{E,B}(l,q) = \frac{\Sigma^2 \, \theta_0^2 }{2} \, e^{-\frac{1}{2}\theta_0^2 (l^2+q^2)}
	\lkw I_0(\theta_0^2 \,l \,q) \pm I_4(\theta_0^2 \,l \,q) \rkw,
	\label{gauss-win1}
\eeq
where $I_0(x)$ and $I_4(x)$ are the modified Bessel functions of the first kind of zeroth and fourth order
respectively \citep{AS}. 
Because of the exponential growth of $I_0(x)$ and $I_4(x)$ with $x$ it is  useful to 
rearrage terms in (\ref{gauss-win1}) and rewrite this equation as  
\beq
	W_{E,B}(l,q) =  \frac{\Sigma^2 \, \theta_0^2}{2} \, e^{-\frac{1}{2}\theta_0^2 (l-q)^2}
	\lkw {\hat I}_0(\theta_0^2 \, l \,q) \pm {\hat I}_4(\theta_0^2 \,l \,q) \rkw,
	\label{gauss-win2}
\eeq
where we have introduced functions ${\hat I}_n(x) = e^{-x} I_n(x)$. 

The large-scale amplitude of $W_{E,B}(l,q)$ can be derived by noting 
the asymptotic behavior of the modified Bessel function for small arguments: $I_0(x) \sim 1$ and 
$I_4(x) \sim x^4/384$ if $x \ll 1$. 
Thus for scales large compared to $R_{1/2}$ when $\theta_0 \,l \ll 1$,
we obtain $W_{E,B}(l, q) \sim \frac{1}{2} \Sigma^2 \, \theta_0^2 \, e^{-\frac{1}{2}\theta_0^2 \,l^2}$.
When we consider power spectra $\Delta P_{E,B}(l)$ in this regime we get the following expression
\beqa
	\Delta P_{E,B}(l) & \approx & \frac{1}{2} \Sigma^2 \theta_0 \int_0^{\infty} dq \, q \, \pkk(q) \, e^{-\frac{1}{2}\theta_0^2 \,q^2} \\
			& \approx & \frac{1}{2} \Sigma^2 \pkk \lok l = \frac{1}{\theta_0} \rok. 
\eeqa
In the above we used the fact that the function $q\, \theta_0 \, e^{-1/2\theta_0^2 \,q^2}$ has a maximum at 
$q = 1/\theta_0$ and can be regarded narrow around its maximum. The error we make using this approximation
is less than $10\%$ for $R_{1/2} = 1\arcdeg$ and $40\%$ for $R_{1/2} = 1\arcmin$. 

For small scales where $\theta_0 \,l \gg 1$ and $\theta_0 \,q \gg 1$,
we may use another  
asymptotic formula for modified Bessel functions which leads to ${\hat I}_n(x) \sim \sqrt{2\pi} \, x^{-1/2}$ \citep{AS}. 
This limit is safely taken when the
argument of ${\hat I}_0$ or ${\hat I}_4$ is greater than $1$ or $100$, respectively. 
Thus we obtain from (\ref{gauss-win2}) for small scales the following 
\beq
\left. 
\begin{array}{l}	
W_{E}(l,q)  \\
W_{B}(l,q) 
\end{array}	
\right\}
\sim  \frac{\Sigma^2 \theta_0^2}{\sqrt{2\pi \theta_0^2 l q}} e^{-\frac{1}{2}\theta_0^2 (l-q)^2} 
\left\{
\begin{array}{l}	
1 - \frac{4}{\theta_0^2 l q}, \\
      \frac{31}{8\theta_0^2 l q}.
\end{array}	
\right.
\label{gauss-win3}
\eeq
The asymptotic expression (\ref{gauss-win3}) is useful when computing 
the small-scale systematics contribution to E and B-mode power spectra (\ref{ps_sys2}). Due to the exponential term 
in the window function (\ref{gauss-win3}) it is effectively a Dirac delta function $\dD(l-q)$.
Thus we can write eq. (\ref{ps_sys2}) as
\beq
\left. 
\begin{array}{l}	
\Delta P_{E}(l) \\ 
\Delta P_{B}(l) 
\end{array}	
\right\}
\sim  \Sigma^2 \, \pkk (l)
\left\{
\begin{array}{l}	
1,  \\
\frac{31}{8} \theta_0^{-2} l^{-2}.
\end{array}	
\right.
\label{gauss-peb}
\eeq
if we consider small scales compared to $R_{1/2}$.
The asymptotic behavior of E and B-mode systematics power spectra is seen in fig. \ref{fig_gauss}.  It is notable that, 
for sufficiently large $l$, the systematic E-mode contribution $\Delta
P_{E}$ is simply a factor
$\Sigma^2$ of the convergence power spectrum $\pkk(l)$. 
Moreover, $\Delta P_{E}$ follows the shape of 
$\pkk(l)$, whereas the B-mode contribution $\Delta P_{B}$ drops rapidly.

\subsection{Patchy}
\label{patchy-type}

Let us consider a systematic signal $\ee(\tht)$ which is perfectly
correlated within circular patches of diameter $\theta_0$ on the
sky. This kind of systematics could arise if the survey is a
mosaic of (circular) telescope pointings, and each pointing has a
constant calibration error that is statistically independent of all
other pointings.  For example, the impact of time-variable
atmospheric seeing on galaxy shape measurements could produce such a pattern.
Assuming such a model, we can compute the correlation function
$\xi^{\ee}(\theta)$ (independent of the specific distribution (pdf) of
$\ee(\tht)$ amplitude)
\beq
	\xi^{\ee}\lok \theta \rok = \Sigma^2 \lkw 1 - \frac{2}{\pi} \lok \arcsin \frac{\theta}{\theta_0} + 
	\frac{\theta}{\theta_0} \sqrt{1-\frac{\theta^2}{\theta_0^2}} \rok \rkw {\mathrm H}(\theta_0 - \theta) 
	\label{linear}
\eeq
where ${\mathrm H}$ is the Heaviside step function and $R_{1/2} \approx 0.404 \theta_0$.  
This type of correlation and its effect on E mode signal degradation and B mode generation
were studied by \cite{2004ApJ...613L...1V} using ray-tracing simulations (their ``sharp modulation'' model). 
Although they assumed
square areas of correlation (the shape of CCD detectors), our analytic
model should match this well if we perform angular averaging over the
square pattern
to get an isotropic correlation function.

In this case a closed form for the window function is not attained,
so we have to rely on numerical integration.
We can deal analytically with the window function (\ref{win}) 
in the limit of small scales $l \theta_0 \gg 1$ and $q \theta_0 \gg 1$. 
For this purpose we can approximate (\ref{linear}) by $\xi^{\ee}\lok \theta \rok \approx \Sigma^2 \lkw 1 - \theta/\theta_0 \rkw$  
where $\theta_0 = 2 R_{1/2}$.  
Let us use asymptotic formulae for the Bessel functions for large
arguments \citep{AS} and write the E mode window function as follows:
\beqa
	W_{E}(l,q) &\approx& \frac{\Sigma^2}{\pi \sqrt{l q}} \nonumber \\
	&& \!\!\!\!\!\!\!\!\!\!\!\!\!\!\!\!\!\!\!\!\!\!\!\!\!\!\!\!\!\!\!\!\! 
	\times \int_0^{\infty} d\theta  
	\lok 1 - \frac{\theta}{\theta_0} \rok \cos \lok l\theta - \frac{\pi}{4} \rok \cos \lok q\theta - \frac{\pi}{4} \rok.
	\label{wlin1}
\eeqa
Using the formulae for addition of cosines
and subsequently perform elementary integration leads us to
\beqa
	W_{E}(l,q) &\approx& \frac{\Sigma^2 \theta_0}{\pi \sqrt{l q}} \nonumber \\
	&& \!\!\!\!\!\!\!\!\!\!\!\!\!\!\!\!\!\!\!\!\!\!\!\!\!\!\!\! 
	\times \lkw \frac{1- \cos \theta_0 \lok l-q \rok}{\theta_0^2 (l-q)^2} + 
	\frac{1- \sin \theta_0 \lok l+q \rok}{\theta_0^2 (l+q)^2} \rkw.
\eeqa
Thus in the interesting case of small scales main contribution to the window function comes from $l \approx q$ which leads to
\beq
	W_{E}(l,q)  \sim  \frac{\Sigma^2 \theta_0}{2 \pi l} 
	\frac{\sin^2 \frac{\theta_0 \lok l-q \rok}{2}}{\frac{\theta_0^2 \lok l-q \rok^2}{4}}
	\sim  \Sigma^2 l^{-1} \dD(l-q).
\eeq
The B-mode window tends to zero because of the identical asymptotic
behavior of $J_0$ and $J_4$. Thus in the small scales limit we have
$\Delta P_{E}(l) \sim \Sigma^2 \, \pkk (l)$, $\Delta P_{B}(l)
\rightarrow 0$ 
which was the case for a Gaussian correlations as well. 

\subsection{Generalized exponential family}
\label{exp-type}

A broad class of correlation functions can be described by a generalized exponential family \citep{Ripley81} 
as follows 
\beq
	\xi^{\ee} (\theta) =\frac{\Sigma^2}{2^{\nu-1} \Gamma(\nu)} \lok \frac{\theta}{\theta_0} \rok^{\nu} 
	K_{\nu} \lok \frac{\theta}{\theta_0} \rok,
	\label{exp}
\eeq
where $K_{\nu}(x)$ is the modified Bessel function of the second kind, 
$\theta_0$ is a characteristic scale  and $0 < \nu < 1$.
For $\nu=1/2$ we obtain an exponential correlation function $\xi^{\ee} (\theta) = \Sigma^2 e^{-\theta/\theta_0}$
with $R_{1/2} = \theta_0 \ln 2$.
When $\nu < 1/2$ the correlation function depends on $\theta$ 
sub-exponentially on small scales and super-exponentially on large scales.
For $\nu > 1/2$ the above behavior is reversed.
Exponential-type correlations decay more slowly than do
Gaussians with the same $R_{1/2}$, so offer a test of the generality of the
behavior of $\Delta P_{E,B}(l)$ for a given characteristic scale
$R_{1/2}$.   The window functions $W_{E,B}$ also decay slowly compared
to the Gaussian case (\S\ref{gauss-type}).

For a generalized exponential family (\ref{exp}) we can compute 
analytically the respective power spectra \citep{GR}
\beq
	P^{\ee}(l) =\frac{ 4 \pi \Sigma^2 \nu}{\theta_0^2}  \lok 1 + \lok l \theta_0 \rok^2 \rok^{-(\nu+1)}.
	\label{exp-ps}
\eeq
The power spectrum has power-law scaling at small scales: $P(l)
\propto l^n$  with $n=-2(\nu+1)$ for $l \theta_0 \gg 1$. The allowed
range of spectral indices is $-4 < n <-2$ (a 
power-law correlation function must have $n > -2$).
An example of a systematic signal of this type could be dust extinction in our Galaxy.
\citet{1998ApJ...500..525S} show that the extinction pattern on the sky
can roughly be described by the power-law power spectrum
$P(l) \propto l^{-5/2}$ for scales larger than $\sim 15\arcmin$,
corresponding to $\nu = 1/4$.

\section{Results}
\label{results}

The cosmological background model we assume is $\Lambda$CDM with
$\Omega_m = 0.3$, $\Omega_{\Lambda}= 1- \Omega_m$, $H_0 = 70\,{\rm
  km}\,{\rm s}^{-1}\,{\rm Mpc}^{-1}$, $\sigma_8 = 0.93$.
The distribution of source galaxies in redshift is assumed to be $dN/dz
\propto z^2 \exp \lkw -\lok z/z_0 \rok^{3/2} \rkw$ with $z_0 = 2/3$
and mean redshift $1$.  These are as assumed by
\citet{2004ApJ...613L...1V}, so that we may test our results against their ray-tracing
simulation results.
We compute the convergence power spectrum $\pkk(l)$ using fitting formula for 3-D dark matter power spectrum given 
by \citet{2003MNRAS.341.1311S}. 

\begin{figure}
\includegraphics[width=8.6cm]{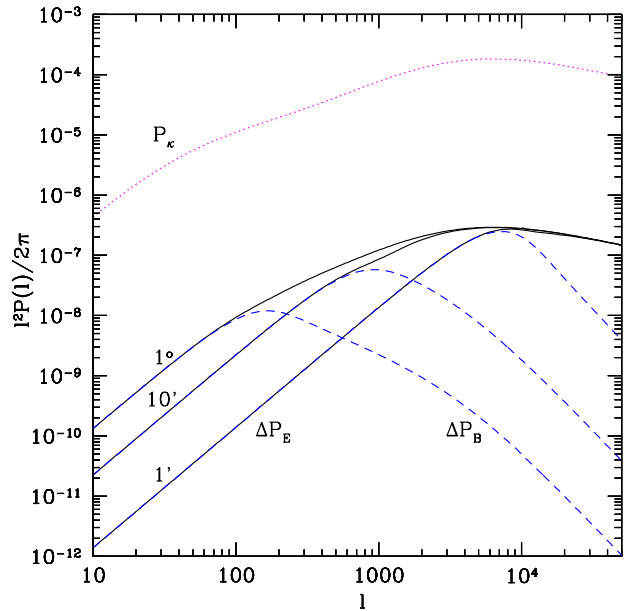}
\caption{\label{fig_gauss} Statistical signal due to multiplicative
  shear errors. Shown are systematic-error contributions to the power spectra in
E mode, $\Delta P_E(l)$, (\emph{solid}) and B mode, $\Delta P_B(l)$, (\emph{dashed}). 
The correlation function $\xie$ of the multiplicative calibration field is assumed to be
  gaussian with $4\%$ rms. 
The three plotted power spectra assume different characteristic scales $R_{1/2}$: $1\arcdeg$, $10\arcmin$ and $1\arcmin$.
At large scales, E mode and B mode contributions are equal to each other for a given characteristic scale. 
Also shown is the convergence power spectrum $\pkk(l)$ (\emph{dotted}).}
\end{figure}

In fig. \ref{fig_gauss} we show the power spectrum contributions due to an
inhomogeneous calibration field described by the Gaussian 
correlation function (\ref{gauss}) for three characteristic scales
$R_{1/2}$: 1\arcdeg, 10\arcmin, and 1\arcmin.
We take the rms of the systematics field to be $\Sigma = 4\%$. 
Recall that $\Delta P_{E,B}(l) \propto \Sigma^2$. 
We notice that $\Delta P_{E}$ spectra are featureless and have maxima
near the maximum $\pkk(l)$ (except for very small
$R_{1/2}$). 
On the other hand, the B-mode power spectra $\Delta P_{B}$ have maxima
near the characteristic scale of the correlation function. 

\begin{figure}
\includegraphics[width=8.6cm]{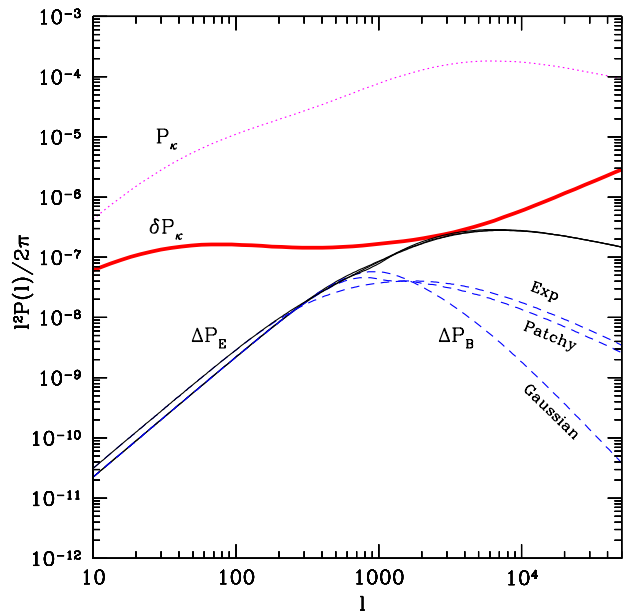}
\caption{\label{fig_stat} Statistical and systematic errors in the
  convergence power spectrum $\pkk(l)$ (dotted): 
the statistical uncertainties $\delta \pkk(l)$ are due to sample variance and source galaxy shape noise (\emph{thick solid}).
The systematic uncertainties in E mode, $\Delta P_E(l)$, (\emph{thin solid}) and B mode, $\Delta P_B(l)$, (\emph{dashed}) 
are shown for different calibration correlation 
models: Gaussian, ``patchy'' and exponential. They all share a
characteristic scale $R_{1/2} = 10\arcmin$ and $\Sigma=4\%$ rms amplitude.
  Note that the E-mode systematic errors are essentially independent
  of the functional form, and are at worst equal to the statistical errors.
}
\end{figure}

In order to assess whether the signal due to systematics can be potentially harmful for weak
lensing results, let us compare the contaminating power spectra $\Delta P_{E,B}(l)$
to the statistical errors on the convergence power spectrum $\delta
\pkk(l)$ \citep{1998ApJ...498...26K}. Assuming gaussianity of the 
convergence field we have  with sufficient accuracy for our purpose that
\beq
	\delta \pkk(l)  = \frac{1}{\left({l\,\Delta l f_{\mathrm{sky}}}\right)^{1/2}} 
	\, \pkk(l) \lok 1 + \frac{\sigma^2_{\gamma}}{n_g \pkk(l)} \rok, 
	\label{rms}
\eeq
where $f_{\mathrm{sky}}$ is the fraction of the sky covered by a survey, $n_g$ is the density of source galaxies 
with measured shapes, and $\sigma_{\gamma}\approx 0.3$ is galaxy shape
noise.  The $\pkk(l)$ data will have to be binned over some
  interval $\Delta l$ for a meaningful comparison with the systematic
  error $\Delta P_{E,B}(l)$.  Because $P_\kappa(l)$ is virtually
  featureless and there is no cosmological information in its detailed
  structure, we choose broad bins of width $\Delta l = l$.  An even
  broader binning scheme would lower the $\delta \pkk(l)$ line in
  Figure~\ref{fig_stat} and our derived requirements
  on $\Sigma$ would scale as $(\Delta l)^{-1/2}$.
Future, ground based, wide-field surveys
like \emph{LSST} \footnote{{\tt http://www.lsst.org}} are expected to cover $f_{\mathrm{sky}} \sim 50\%$ of the sky
and obtain good shape measurements for about $30$ galaxies per $\mathrm{arcmin}^{2}$. 
Figure~\ref{fig_stat} shows the convergence power spectrum $\pkk(l)$ and its statistical errors (\ref{rms})  
for these values of $f_{\mathrm{sky}}$ and $n_g$. The encouraging implication of fig. \ref{fig_stat} is that keeping 
systematics (e.g, shear calibration errors) below $3\%$  rms 
($\Sigma\lesssim 3\%$) should avoid significant contamination of the
observed $\pkk(l)$, even for future surveys.       

Figure~\ref{fig_stat} plots systematic-error power spectra $\Delta P_{E,B}(l)$ for a gaussian (\ref{gauss}), ``patchy'' (\ref{linear}), 
and exponential (\ref{exp}) correlation functions, each with
$R_{1/2}=10\arcmin$ and $\Sigma=4\%$.
We notice that the shape of $\Delta P_{E,B}(l)$ is nearly independent of
the specific shape of the correlation function $\ee(\tht)$ 
of the systematic field.
Thus from a practical point of view the important features of the systematic field are the characteristic scale 
of correlations and the rms of the field.  The latter affects the
overall amplitude of the systematic errors as $\Delta P_{E,B}(l) \propto\Sigma^2$.
The former fixes the amplitude of the $\Delta P_E(l)=\Delta P_B(l)$ at
large scales, and gives the scale where the B mode starts decaying.  

\begin{figure}
\includegraphics[width=8.6cm]{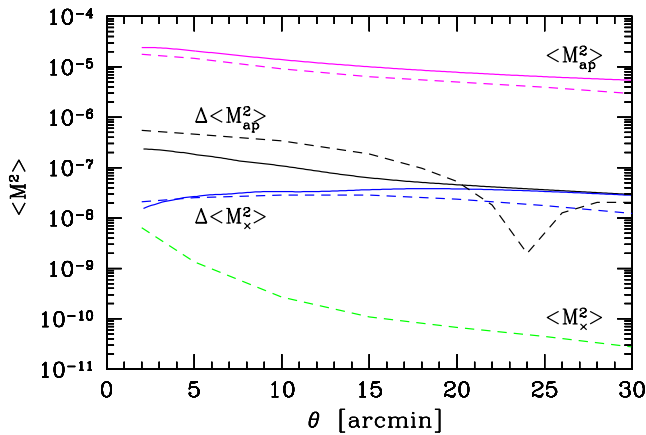}
\caption{\label{fig_mass} Aperture mass variances as computed
  analytically in this work (\emph{solid}) and obtained from 
ray tracing simulations \citep{2004ApJ...613L...1V} (\emph{dashed}): there is a good agreement between 
these two approaches. The two upper curves, denoted by
$\lensa M_{{\rm ap}}^2 \rensa$ represent the signal from the cosmological E mode (without systematics) and
the lowest curve, $\lensa M_{{\times}}^2 \rensa$, shows the cosmological contribution to B mode from 
simulations (in our case this contribution is zero). The two pairs of curves denoted  $\Delta\lensa M_{{\rm ap}}^2 \rensa$
and $\Delta\lensa M_{{\times}}^2 \rensa$
show the systematic signal contribution to the aperture mass variance in E mode 
and B mode, respectively.  
The underlying modulation is of ``patchy'' type with $\theta_0 = 25\arcmin$ and $10\%$ rms (sec. \ref{patchy-type}).}
\end{figure}

We can compare our analytic results for the ``patchy'' correlation function to
the numerical tests of 
\citet{2004ApJ...613L...1V}.  We
set $\theta_0 = 25\arcmin$ and $\Sigma = 10\%$ to match the
calibration-error pattern they superpose on their ray-tracing data.
Our analytic estimates
of the errors induced in the
aperture mass variances $\lensa M_{{\rm ap}, \times}^2 \rensa (R)$ 
are shown in fig.~\ref{fig_mass}. 
These errors can be directly compared to those shown in Figure~2 of
\citet{2004ApJ...613L...1V}, which we reproduce in our Figure.  
Our estimates closely reproduce the results of the numerical simulation,
except that we do not produce trough of $\Delta \lensa M_{{\rm ap}}^2
\rensa$ at the characteristic scale around  
$25 \, \mathrm{arcmin}$.  The trough might be attributable to sample variance
from the finite number (64) of patches used in the simulated images.

\section{Conclusions}

We consider the effect of spatially varying multiplicative systematic errors (assumed
uncorrelated with the cosmological signal) on the measured power
spectra $P_E(l)$ and $P_B(l)$ in the case of the 2D lensing. The prime example
of this type of systematic would be shear calibration errors which
vary across the survey area due to changing observing conditions.
As shown by \citet{2003MNRAS.343..459H} overall shear calibration errors of existing methods of shear measurement
can reach $10\%$ for galaxies of size comparable to the PSF.  Such errors
grow larger if one uses more poorly-resolved galaxies, as
would be the case for deep ground-based surveys like \emph{LSST}. 
Uncorrected Galactic extinction could also introduce spatially correlated
systematics in survey depth, altering the observed shear correlation functions.
When we examine a variety of functional forms for the correlation
function $\xie$ of the inhomogeneous systematic, we find that all salient
effects on the measured power spectrum can be
 characterized by the variance $\Sigma^2$
of the correlation and its characteristic scale $R_{1/2}$.
A wide variety of functional forms for $\xie$ induced very similar
effects on measurements of the convergence power spectrum. Only the
small-scale B-mode spectrum is sensitive to the detailed shape of $\xie$.

Comparison of the systematics errors $\Delta P_E(l)$ on the power
spectrum to the statistical errors expected for future 
weak lensing surveys indicates that we should not be afraid of
systematic contamination if we keep calibration errors below
$\Sigma\approx3\%$.

The absence of B-mode contamination in the most recent cosmic-shear
measurements 
\citep{2005A&A...429...75V, 2004astro.ph.12234J, 2005astro.ph..1201M}
suggests that systematic errors of the type considered here are below
current  statistical errors (5--10\%) and
hence do not bias the conclusions.
Note, however, that B-mode power $\Delta P_B(l)$ consistent with zero 
on scales $l \gg 1/\theta_0$ does not necessarily imply the absence of
significant calibration error $\Delta P_E(l)$ (see Figure~
\ref{fig_gauss}). 
Hence the present cosmic shear results could be significantly affected
by calibration errors, if they
have a correlation length $\theta_0$ that is larger than scales
considered 
in the B-mode measurement ($\theta_0 \gg 1/l$).
Future surveys will beat
down statistical errors, so we will have to understand and beat down
systematic errors as well.  This work suggests that spatially-varying
calibration errors will have to be reduced to 3\%.  This is well below
the levels that have been demonstrated to date, but is probably
achievable for well-behaved data with careful shape-measurement
techniques \citep{2005astro.ph..6112H, Reiko2005}.

\begin{acknowledgments}
We would like to thank Bhuvnesh Jain for frequent discussions. 
J.G. would like to thank Laura Marian for help with
\textsc{Mathematica}.
This work is supported by grants AST-0236702 from the National
Science Foundation, Department of Energy grant
DOE-DE-FG02-95ER40893, and Polish State Committee for Scientific 
Research grant 1P03D01226. 
\end{acknowledgments}

%---------------------------------------------------------------------------------------------
\def\apjl{Astrophys.\ J.\ Lett.}
\def\mnras{Mon.\ Not.\ R.\ Astron.\ Soc.}
\def\nat{Nature (London)}
\def\araa{Ann. Rev. Astron. Astrophys.}
\def\aap{Astron.\ Astrophys.}
\def\apj{Astrophys.\ J.}
\def\aj{Astron.\ J.}
\def\apjs{Astrophys.\ J. Supp.}
\def\prl{Phys.\ Rev.\ Lett.}
\def\prd{Phys.\ Rev.\ D}
\def\pl{Phys.\ Lett.}
\def\physrep{Phys. Rep.}
%---------------------------------------------------------------------------------------------


\begin{thebibliography}{30}
\expandafter\ifx\csname natexlab\endcsname\relax\def\natexlab#1{#1}\fi
\expandafter\ifx\csname bibnamefont\endcsname\relax
  \def\bibnamefont#1{#1}\fi
\expandafter\ifx\csname bibfnamefont\endcsname\relax
  \def\bibfnamefont#1{#1}\fi
\expandafter\ifx\csname citenamefont\endcsname\relax
  \def\citenamefont#1{#1}\fi
\expandafter\ifx\csname url\endcsname\relax
  \def\url#1{\texttt{#1}}\fi
\expandafter\ifx\csname urlprefix\endcsname\relax\def\urlprefix{URL }\fi
\providecommand{\bibinfo}[2]{#2}
\providecommand{\eprint}[2][]{\url{#2}}

\bibitem[{\citenamefont{{Miralda-Escude}}(1991)}]{1991ApJ...380....1M}
\bibinfo{author}{\bibfnamefont{J.}~\bibnamefont{{Miralda-Escude}}},
  \bibinfo{journal}{\apj} \textbf{\bibinfo{volume}{380}}, \bibinfo{pages}{1}
  (\bibinfo{year}{1991}).

\bibitem[{\citenamefont{{Kaiser}}(1992)}]{1992ApJ...388..272K}
\bibinfo{author}{\bibfnamefont{N.}~\bibnamefont{{Kaiser}}},
  \bibinfo{journal}{\apj} \textbf{\bibinfo{volume}{388}}, \bibinfo{pages}{272}
  (\bibinfo{year}{1992}).

\bibitem[{\citenamefont{{Jarvis} et~al.}(2005)\citenamefont{{Jarvis}, {Jain},
  {Bernstein}, and {Dolney}}}]{2005astro.ph..2243J}
\bibinfo{author}{\bibfnamefont{M.}~\bibnamefont{{Jarvis}}},
  \bibinfo{author}{\bibfnamefont{B.}~\bibnamefont{{Jain}}},
  \bibinfo{author}{\bibfnamefont{G.}~\bibnamefont{{Bernstein}}},
  \bibnamefont{and} \bibinfo{author}{\bibfnamefont{D.}~\bibnamefont{{Dolney}}},
  \bibinfo{journal}{astro-ph/0502243}  (\bibinfo{year}{2005}).

\bibitem[{\citenamefont{{Huterer} et~al.}(2005)\citenamefont{{Huterer},
  {Takada}, {Bernstein}, and {Jain}}}]{2005astro.ph..6030H}
\bibinfo{author}{\bibfnamefont{D.}~\bibnamefont{{Huterer}}},
  \bibinfo{author}{\bibfnamefont{M.}~\bibnamefont{{Takada}}},
  \bibinfo{author}{\bibfnamefont{G.}~\bibnamefont{{Bernstein}}},
  \bibnamefont{and} \bibinfo{author}{\bibfnamefont{B.}~\bibnamefont{{Jain}}},
  \bibinfo{journal}{astro-ph/0506030}  (\bibinfo{year}{2005}).

\bibitem[{\citenamefont{{Bernstein} and {Jarvis}}(2002)}]{2002AJ....123..583B}
\bibinfo{author}{\bibfnamefont{G.~M.} \bibnamefont{{Bernstein}}}
  \bibnamefont{and} \bibinfo{author}{\bibfnamefont{M.}~\bibnamefont{{Jarvis}}},
  \bibinfo{journal}{\aj} \textbf{\bibinfo{volume}{123}}, \bibinfo{pages}{583}
  (\bibinfo{year}{2002}).

\bibitem[{\citenamefont{{Jarvis} et~al.}(2003)\citenamefont{{Jarvis},
  {Bernstein}, {Fischer}, {Smith}, {Jain}, {Tyson}, and
  {Wittman}}}]{2003AJ....125.1014J}
\bibinfo{author}{\bibfnamefont{M.}~\bibnamefont{{Jarvis}}},
  \bibinfo{author}{\bibfnamefont{G.~M.} \bibnamefont{{Bernstein}}},
  \bibinfo{author}{\bibfnamefont{P.}~\bibnamefont{{Fischer}}},
  \bibinfo{author}{\bibfnamefont{D.}~\bibnamefont{{Smith}}},
  \bibinfo{author}{\bibfnamefont{B.}~\bibnamefont{{Jain}}},
  \bibinfo{author}{\bibfnamefont{J.~A.} \bibnamefont{{Tyson}}},
  \bibnamefont{and}
  \bibinfo{author}{\bibfnamefont{D.}~\bibnamefont{{Wittman}}},
  \bibinfo{journal}{\aj} \textbf{\bibinfo{volume}{125}}, \bibinfo{pages}{1014}
  (\bibinfo{year}{2003}).

\bibitem[{\citenamefont{{Schneider} et~al.}(2002)\citenamefont{{Schneider},
  {van Waerbeke}, and {Mellier}}}]{2002A&A...389..729S}
\bibinfo{author}{\bibfnamefont{P.}~\bibnamefont{{Schneider}}},
  \bibinfo{author}{\bibfnamefont{L.}~\bibnamefont{{van Waerbeke}}},
  \bibnamefont{and}
  \bibinfo{author}{\bibfnamefont{Y.}~\bibnamefont{{Mellier}}},
  \bibinfo{journal}{\aap} \textbf{\bibinfo{volume}{389}}, \bibinfo{pages}{729}
  (\bibinfo{year}{2002}).

\bibitem[{\citenamefont{{Vale} et~al.}(2004)\citenamefont{{Vale}, {Hoekstra},
  {van Waerbeke}, and {White}}}]{2004ApJ...613L...1V}
\bibinfo{author}{\bibfnamefont{C.}~\bibnamefont{{Vale}}},
  \bibinfo{author}{\bibfnamefont{H.}~\bibnamefont{{Hoekstra}}},
  \bibinfo{author}{\bibfnamefont{L.}~\bibnamefont{{van Waerbeke}}},
  \bibnamefont{and} \bibinfo{author}{\bibfnamefont{M.}~\bibnamefont{{White}}},
  \bibinfo{journal}{\apjl} \textbf{\bibinfo{volume}{613}}, \bibinfo{pages}{L1}
  (\bibinfo{year}{2004}).

\bibitem[{\citenamefont{{Stebbins}}(1996)}]{1996astro.ph..9149S}
\bibinfo{author}{\bibfnamefont{A.}~\bibnamefont{{Stebbins}}},
  \bibinfo{journal}{astro-ph/9609149}  (\bibinfo{year}{1996}).

\bibitem[{\citenamefont{{Kamionkowski}
  et~al.}(1997)\citenamefont{{Kamionkowski}, {Kosowsky}, and
  {Stebbins}}}]{1997PhRvD..55.7368K}
\bibinfo{author}{\bibfnamefont{M.}~\bibnamefont{{Kamionkowski}}},
  \bibinfo{author}{\bibfnamefont{A.}~\bibnamefont{{Kosowsky}}},
  \bibnamefont{and}
  \bibinfo{author}{\bibfnamefont{A.}~\bibnamefont{{Stebbins}}},
  \bibinfo{journal}{\prd} \textbf{\bibinfo{volume}{55}}, \bibinfo{pages}{7368}
  (\bibinfo{year}{1997}).

\bibitem[{\citenamefont{{Zaldarriaga} and
  {Seljak}}(1997)}]{1997PhRvD..55.1830Z}
\bibinfo{author}{\bibfnamefont{M.}~\bibnamefont{{Zaldarriaga}}}
  \bibnamefont{and} \bibinfo{author}{\bibfnamefont{U.}~\bibnamefont{{Seljak}}},
  \bibinfo{journal}{\prd} \textbf{\bibinfo{volume}{55}}, \bibinfo{pages}{1830}
  (\bibinfo{year}{1997}).

\bibitem[{\citenamefont{{Crittenden} et~al.}(2002)\citenamefont{{Crittenden},
  {Natarajan}, {Pen}, and {Theuns}}}]{2002ApJ...568...20C}
\bibinfo{author}{\bibfnamefont{R.~G.} \bibnamefont{{Crittenden}}},
  \bibinfo{author}{\bibfnamefont{P.}~\bibnamefont{{Natarajan}}},
  \bibinfo{author}{\bibfnamefont{U.}~\bibnamefont{{Pen}}}, \bibnamefont{and}
  \bibinfo{author}{\bibfnamefont{T.}~\bibnamefont{{Theuns}}},
  \bibinfo{journal}{\apj} \textbf{\bibinfo{volume}{568}}, \bibinfo{pages}{20}
  (\bibinfo{year}{2002}).

\bibitem[{\citenamefont{{Bartelmann} and
  {Schneider}}(2001)}]{2001PhR...340..291B}
\bibinfo{author}{\bibfnamefont{M.}~\bibnamefont{{Bartelmann}}}
  \bibnamefont{and}
  \bibinfo{author}{\bibfnamefont{P.}~\bibnamefont{{Schneider}}},
  \bibinfo{journal}{\physrep} \textbf{\bibinfo{volume}{340}},
  \bibinfo{pages}{291} (\bibinfo{year}{2001}).

\bibitem[{\citenamefont{{Van Waerbeke} and
  {Mellier}}(2003)}]{2003astro.ph..5089V}
\bibinfo{author}{\bibfnamefont{L.}~\bibnamefont{{Van Waerbeke}}}
  \bibnamefont{and}
  \bibinfo{author}{\bibfnamefont{Y.}~\bibnamefont{{Mellier}}},
  \bibinfo{journal}{astro-ph/0305089}  (\bibinfo{year}{2003}).

\bibitem[{\citenamefont{{Hirata} and {Seljak}}(2003)}]{2003MNRAS.343..459H}
\bibinfo{author}{\bibfnamefont{C.}~\bibnamefont{{Hirata}}} \bibnamefont{and}
  \bibinfo{author}{\bibfnamefont{U.}~\bibnamefont{{Seljak}}},
  \bibinfo{journal}{\mnras} \textbf{\bibinfo{volume}{343}},
  \bibinfo{pages}{459} (\bibinfo{year}{2003}).

\bibitem[{\citenamefont{{Van Waerbeke} et~al.}(2005)\citenamefont{{Van
  Waerbeke}, {Mellier}, and {Hoekstra}}}]{2005A&A...429...75V}
\bibinfo{author}{\bibfnamefont{L.}~\bibnamefont{{Van Waerbeke}}},
  \bibinfo{author}{\bibfnamefont{Y.}~\bibnamefont{{Mellier}}},
  \bibnamefont{and}
  \bibinfo{author}{\bibfnamefont{H.}~\bibnamefont{{Hoekstra}}},
  \bibinfo{journal}{\aap} \textbf{\bibinfo{volume}{429}}, \bibinfo{pages}{75}
  (\bibinfo{year}{2005}).

\bibitem[{\citenamefont{{Jarvis} and {Jain}}(2004)}]{2004astro.ph.12234J}
\bibinfo{author}{\bibfnamefont{M.}~\bibnamefont{{Jarvis}}} \bibnamefont{and}
  \bibinfo{author}{\bibfnamefont{B.}~\bibnamefont{{Jain}}},
  \bibinfo{journal}{astro-ph/0412234}  (\bibinfo{year}{2004}).

\bibitem[{\citenamefont{{Heymans} and {Heavens}}(2003)}]{2003MNRAS.339..711H}
\bibinfo{author}{\bibfnamefont{C.}~\bibnamefont{{Heymans}}} \bibnamefont{and}
  \bibinfo{author}{\bibfnamefont{A.}~\bibnamefont{{Heavens}}},
  \bibinfo{journal}{\mnras} \textbf{\bibinfo{volume}{339}},
  \bibinfo{pages}{711} (\bibinfo{year}{2003}).

\bibitem[{\citenamefont{{Kaiser} et~al.}(1995)\citenamefont{{Kaiser},
  {Squires}, and {Broadhurst}}}]{1995ApJ...449..460K}
\bibinfo{author}{\bibfnamefont{N.}~\bibnamefont{{Kaiser}}},
  \bibinfo{author}{\bibfnamefont{G.}~\bibnamefont{{Squires}}},
  \bibnamefont{and}
  \bibinfo{author}{\bibfnamefont{T.}~\bibnamefont{{Broadhurst}}},
  \bibinfo{journal}{\apj} \textbf{\bibinfo{volume}{449}}, \bibinfo{pages}{460}
  (\bibinfo{year}{1995}).

\bibitem[{\citenamefont{{Abramowitz} and {Stegun}}(1965)}]{AS}
\bibinfo{author}{\bibfnamefont{M.}~\bibnamefont{{Abramowitz}}}
  \bibnamefont{and} \bibinfo{author}{\bibfnamefont{I.}~\bibnamefont{{Stegun}}},
  \emph{\bibinfo{title}{Handbook of Mathematical Functions}}
  (\bibinfo{publisher}{Dover Publications}, \bibinfo{address}{New York},
  \bibinfo{year}{1965}).

\bibitem[{\citenamefont{{Schneider} et~al.}(1998)\citenamefont{{Schneider},
  {van Waerbeke}, {Jain}, and {Kruse}}}]{1998MNRAS.296..873S}
\bibinfo{author}{\bibfnamefont{P.}~\bibnamefont{{Schneider}}},
  \bibinfo{author}{\bibfnamefont{L.}~\bibnamefont{{van Waerbeke}}},
  \bibinfo{author}{\bibfnamefont{B.}~\bibnamefont{{Jain}}}, \bibnamefont{and}
  \bibinfo{author}{\bibfnamefont{G.}~\bibnamefont{{Kruse}}},
  \bibinfo{journal}{\mnras} \textbf{\bibinfo{volume}{296}},
  \bibinfo{pages}{873} (\bibinfo{year}{1998}).

\bibitem[{\citenamefont{{Bartelmann} and
  {Schneider}}(1999)}]{1999A&A...345...17B}
\bibinfo{author}{\bibfnamefont{M.}~\bibnamefont{{Bartelmann}}}
  \bibnamefont{and}
  \bibinfo{author}{\bibfnamefont{P.}~\bibnamefont{{Schneider}}},
  \bibinfo{journal}{\aap} \textbf{\bibinfo{volume}{345}}, \bibinfo{pages}{17}
  (\bibinfo{year}{1999}).

\bibitem[{\citenamefont{{Ripley}}(1981)}]{Ripley81}
\bibinfo{author}{\bibfnamefont{B.}~\bibnamefont{{Ripley}}},
  \emph{\bibinfo{title}{Spatial Statistics}} (\bibinfo{publisher}{John Wiley
  and Sons}, \bibinfo{address}{New York}, \bibinfo{year}{1981}).

\bibitem[{\citenamefont{{Gradshteyn} and {Ryzhik}}(2000)}]{GR}
\bibinfo{author}{\bibfnamefont{I.}~\bibnamefont{{Gradshteyn}}}
  \bibnamefont{and} \bibinfo{author}{\bibfnamefont{I.}~\bibnamefont{{Ryzhik}}},
  \emph{\bibinfo{title}{Table of Integrals, Series and Products. Sixth
  Edition}} (\bibinfo{publisher}{Academic Press}, \bibinfo{address}{San Diego},
  \bibinfo{year}{2000}).

\bibitem[{\citenamefont{{Schlegel} et~al.}(1998)\citenamefont{{Schlegel},
  {Finkbeiner}, and {Davis}}}]{1998ApJ...500..525S}
\bibinfo{author}{\bibfnamefont{D.~J.} \bibnamefont{{Schlegel}}},
  \bibinfo{author}{\bibfnamefont{D.~P.} \bibnamefont{{Finkbeiner}}},
  \bibnamefont{and} \bibinfo{author}{\bibfnamefont{M.}~\bibnamefont{{Davis}}},
  \bibinfo{journal}{\apj} \textbf{\bibinfo{volume}{500}}, \bibinfo{pages}{525}
  (\bibinfo{year}{1998}).

\bibitem[{\citenamefont{{Smith} et~al.}(2003)\citenamefont{{Smith}, {Peacock},
  {Jenkins}, {White}, {Frenk}, {Pearce}, {Thomas}, {Efstathiou}, and
  {Couchman}}}]{2003MNRAS.341.1311S}
\bibinfo{author}{\bibfnamefont{R.~E.} \bibnamefont{{Smith}}},
  \bibinfo{author}{\bibfnamefont{J.~A.} \bibnamefont{{Peacock}}},
  \bibinfo{author}{\bibfnamefont{A.}~\bibnamefont{{Jenkins}}},
  \bibinfo{author}{\bibfnamefont{S.~D.~M.} \bibnamefont{{White}}},
  \bibinfo{author}{\bibfnamefont{C.~S.} \bibnamefont{{Frenk}}},
  \bibinfo{author}{\bibfnamefont{F.~R.} \bibnamefont{{Pearce}}},
  \bibinfo{author}{\bibfnamefont{P.~A.} \bibnamefont{{Thomas}}},
  \bibinfo{author}{\bibfnamefont{G.}~\bibnamefont{{Efstathiou}}},
  \bibnamefont{and} \bibinfo{author}{\bibfnamefont{H.~M.~P.}
  \bibnamefont{{Couchman}}}, \bibinfo{journal}{\mnras}
  \textbf{\bibinfo{volume}{341}}, \bibinfo{pages}{1311} (\bibinfo{year}{2003}).

\bibitem[{\citenamefont{{Kaiser}}(1998)}]{1998ApJ...498...26K}
\bibinfo{author}{\bibfnamefont{N.}~\bibnamefont{{Kaiser}}},
  \bibinfo{journal}{\apj} \textbf{\bibinfo{volume}{498}}, \bibinfo{pages}{26}
  (\bibinfo{year}{1998}).

\bibitem[{\citenamefont{{Mandelbaum} et~al.}(2005)\citenamefont{{Mandelbaum},
  {Hirata}, {Seljak}, {Guzik}, {Padmanabhan}, {Blake}, {Blanton}, {Lupton}, and
  {Brinkmann}}}]{2005astro.ph..1201M}
\bibinfo{author}{\bibfnamefont{R.}~\bibnamefont{{Mandelbaum}}},
  \bibinfo{author}{\bibfnamefont{C.~M.} \bibnamefont{{Hirata}}},
  \bibinfo{author}{\bibfnamefont{U.}~\bibnamefont{{Seljak}}},
  \bibinfo{author}{\bibfnamefont{J.}~\bibnamefont{{Guzik}}},
  \bibinfo{author}{\bibfnamefont{N.}~\bibnamefont{{Padmanabhan}}},
  \bibinfo{author}{\bibfnamefont{C.}~\bibnamefont{{Blake}}},
  \bibinfo{author}{\bibfnamefont{M.~R.} \bibnamefont{{Blanton}}},
  \bibinfo{author}{\bibfnamefont{R.}~\bibnamefont{{Lupton}}}, \bibnamefont{and}
  \bibinfo{author}{\bibfnamefont{J.}~\bibnamefont{{Brinkmann}}},
  \bibinfo{journal}{astro-ph/0501201}  (\bibinfo{year}{2005}).

\bibitem[{\citenamefont{{Heymans} et~al.}(2005)\citenamefont{{Heymans}, {Van
  Waerbeke}, {Bacon}, {Berge}, {Bernstein}, {Bertin}, {Bridle}, {Brown},
  {Clowe}, {Dahle} et~al.}}]{2005astro.ph..6112H}
\bibinfo{author}{\bibfnamefont{C.}~\bibnamefont{{Heymans}}},
  \bibinfo{author}{\bibfnamefont{L.}~\bibnamefont{{Van Waerbeke}}},
  \bibinfo{author}{\bibfnamefont{D.}~\bibnamefont{{Bacon}}},
  \bibinfo{author}{\bibfnamefont{J.}~\bibnamefont{{Berge}}},
  \bibinfo{author}{\bibfnamefont{G.}~\bibnamefont{{Bernstein}}},
  \bibinfo{author}{\bibfnamefont{E.}~\bibnamefont{{Bertin}}},
  \bibinfo{author}{\bibfnamefont{S.}~\bibnamefont{{Bridle}}},
  \bibinfo{author}{\bibfnamefont{M.~L.} \bibnamefont{{Brown}}},
  \bibinfo{author}{\bibfnamefont{D.}~\bibnamefont{{Clowe}}},
  \bibinfo{author}{\bibfnamefont{H.}~\bibnamefont{{Dahle}}},
  \bibnamefont{et~al.}, \bibinfo{journal}{astro-ph/0506112}
  (\bibinfo{year}{2005}).

\bibitem[{\citenamefont{{Nakajima} and {Bernstein}}(2005)}]{Reiko2005}
\bibinfo{author}{\bibfnamefont{R.}~\bibnamefont{{Nakajima}}} \bibnamefont{and}
  \bibinfo{author}{\bibfnamefont{G.}~\bibnamefont{{Bernstein}}},
  \bibinfo{journal}{in preparation}  (\bibinfo{year}{2005}).

\end{thebibliography}
\end{document}